\documentclass{aa}
\usepackage{psfig}

\def\la{\;
\raise0.3ex\hbox{$<$\kern-0.75em\raise-1.1ex\hbox{$\sim$}}\; }
\def\ga{\;
\raise0.3ex\hbox{$>$\kern-0.75em\raise-1.1ex\hbox{$\sim$}}\; }

\newcommand{\zabs}{$z_{\rm abs}\,$}
\newcommand{\zem}{$z_{\rm em}\,$}

\newcommand{\kms}{km~s$^{-1}\,$}
\newcommand{\ms}{m~s$^{-1}\,$}
\newcommand{\cm}{cm$^{-2}\,$}

\newcommand{\daa}{$\Delta\alpha/\alpha\,$}

\begin{document}

\title{Most precise single redshift bound to 
$\Delta\alpha/\alpha$\thanks{Based on 
observations performed at the VLT Kueyen telescope (ESO, Paranal, Chile). 
The data are retrieved  from the ESO/ST-ECF Science Archive Facility.
}
}
\author{
S. A. Levshakov\inst{1}
\and
M. Centuri\'on\inst{2}
\and
P. Molaro\inst{2,3}
\and
S. D'Odorico\inst{4}
\and\\
D. Reimers\inst{5}
\and
R. Quast\inst{5}
\and
M. Pollmann\inst{5}
}
\offprints{S.~A.~Levshakov
\protect \\lev@astro.ioffe.rssi.ru}
\institute{
Department of Theoretical Astrophysics,
Ioffe Physico-Technical Institute, 194021 St.Petersburg, Russia
\and
Osservatorio Astronomico di Trieste, Via G. B. Tiepolo 11,
34131 Trieste, Italy
\and
Observatoire de Paris 61, avenue de l'Observatoire, 75014 Paris, France
\and
European Southern Observatory, Karl-Schwarzschild-Strasse 2,
D-85748 Garching bei M\"unchen, Germany
\and
Hamburger Sternwarte, Universit\"at Hamburg,
Gojenbergsweg 112, D-21029 Hamburg, Germany
}
\date{Received 00  / Accepted 00 }
\abstract{Verification of
theoretical predictions of an oscillating behavior of the
fine-structure constant $\alpha\,\, (\equiv e^2/\hbar c)$
with cosmic time requires high precision \daa\ measurements at 
individual redshifts, while in earlier studies the mean \daa\ values 
averaged over wide redshift intervals were usually reported. 
This requirement can be met via
the single ion differential $\alpha$ measurement (SIDAM) procedure
proposed in Levshakov et al. (2005). 
We apply the SIDAM to the \ion{Fe}{ii} lines 
associated with the damped Ly$\alpha$
system observed at \zabs = 1.15 in the spectrum of \object{HE 0515--4414}. 
The weighted mean 
$\langle \Delta\alpha/\alpha \rangle$ calculated on base of carefully 
selected 34 
\ion{Fe}{ii} pairs $\{\lambda 1608,X\}$ ($X = 2344, 2374$, and 2586 \AA)
is 
$\langle \Delta\alpha/\alpha \rangle =
(-0.07\pm0.84)\times10^{-6}$ ($1\sigma$ C.L.).
The precision of this estimate
improves by a factor 2 the previous one reported for the same system by
Quast et al. (2004).
The obtained result represents an absolute improvement with
respect to what has been done in the measurements of \daa.
\keywords{Cosmology: observations -- Line: profiles -- 
Quasars: absorption lines --
Quasars: individual: \object{HE 0515--4414}}
} 
\authorrunning{S. A. Levshakov et al.}
\titlerunning{Most precise single redshift bound to $\Delta\alpha/\alpha$}
\maketitle

\section{Introduction}

One of the most important physical constants, the
Sommerfeld fine-structure constant $\alpha$, is subjected 
to more and more precise measurements in  both
modern laboratory experiments
and astronomical observations due to fundamental r\^ole which
$\alpha$ plays
in quantum electrodynamics effects and
in electromagnetic and optical properties of atoms.
An unprecedented accuracy of $\sim 0.1\times10^{-9}$ 
(the relative error)
has been recently achieved in the laboratory measurements of the atomic
transitions in alkali atoms (Banerjee et al. 2004). 
This result, combined with sub-ppb (parts per billion) measurements
of the atomic masses (Bradley et al. 1999), enables a high-precision
determination of the fine-structure constant at the present
time, i.e. at the cosmological epoch $z=0$.

%-------------------- Table 1
\begin{table*}[t]
\centering
\caption{ESO UVES archive data on the quasar \object{HE 0515--4414} }
\label{tbl-1}
\begin{tabular}{ccccccccccc}
\hline
\noalign{\smallskip}
\multicolumn{1}{l}{Exp.} & \multicolumn{1}{l}{Set-} & \multicolumn{1}{l}{Slit,} & 
Date, & Time, & Exposure, & 
\multicolumn{1}{l}{Seeing, $\cal{S}$,} & 
\multicolumn{1}{c}{T, $^\circ$C}  & 
\multicolumn{1}{c}{P, mb } & Quality, & Programme \\ 
\multicolumn{1}{l}{No.} & \multicolumn{1}{l}{ting} & 
\multicolumn{1}{l}{arcsec} & y-m-d & h:m & h:m & 
\multicolumn{1}{l}{arcsec} & & & $\kappa$ & ID \\ 
(1) & (2) & (3) & (4) & (5) & (6) & (7) & 
(8) & (9) & (10) & (11) \\
\hline
\noalign{\smallskip}
\noalign{\smallskip}
1b&346&0.8&1999-12-14&05:27&1:15&0.46--0.56&11.8--12.8&990.4--991.0&
0.2&60.A-9022\\ 
1r&580&0.7& &05:27& & & &  & & \\ 
\noalign{\smallskip}
2b&346&0.8&1999-12-14&06:43&1:15&0.56--1.04&11.5--11.9&990.1--990.3&
0.3&60.A-9022\\ 
2r&580&0.7& &06:43& & & &  & & \\ 
\noalign{\smallskip}
3b&346&0.8&1999-12-15&04:47&1:30&0.51--1.84&10.1--11.9&990.0--990.8&
0.006&60.A-9022\\ 
3r&580&0.7& &04:47& & & &  & & \\ 
\noalign{\smallskip}
4b&346&0.8&2000-11-17&06:42&1:15&0.34--0.53&13.5--13.8&990.1--990.3&
1.0&66.A-0212\\ 
4r&580&0.8& &06:42& & & &  & & \\ 
\noalign{\smallskip}
5b&346&0.8&2000-11-18&07:25&1:15&0.55--0.85&13.2--13.7&989.1--989.2&
0.8&66.A-0212\\ 
5r&580&0.8& &07:25& & & &  & & \\ 
\noalign{\smallskip}
6b&346&0.8&2000-11-19&07:27&1:15&0.58--0.80&13.2--14.0&989.2--989.5&
0.2&66.A-0212\\ 
6r&580&0.8& &07:27& & & &  & & \\ 
\noalign{\smallskip}
7b&346&0.8&2000-12-25&06:13&1:15&0.55--0.88&15.8--16.5&989.9--990.0&
0.5&66.A-0212\\ 
7r&580&0.8& &06:13& & & &  & & \\ 
\noalign{\smallskip}
\hline
\noalign{\smallskip}
\multicolumn{11}{l}{Seeing, ${\cal S}$, temperature,
$T$, and air pressure, $P$, limiting values 
listed, respectively, in Cols.7, 8 and 9 are taken from }\\ 
\multicolumn{11}{l}{the ESO/ST-ECF Science Archive Ambient Conditions Database 
at\, http://archive.eso.org/}\\ 
\end{tabular}
\end{table*}

The question whether or not the value of $\alpha$ varied at different
cosmological epochs can be answered only through spectral studies
of extragalactic objects. 
Theoretically the effects of inhomogeneous space and time evolution
of $\alpha$ were considered by Marciano (1984) and Mota \& Barrow (2004).   
Most recently Fujii \& Mizuno (2005) and Fujii (2005) suggested
a damped-oscillation-like behavior of $\alpha$ as a function of 
cosmic time $t$.
It is apparent that to
study such irregular changes in $\alpha$,
we need to achieve high precision in the measurements of
\daa\ at {\it individual} redshifts,
contrary to the
averaging procedure over many redshifts 
which is usually used to decrease uncertainties of the mean values 
$\langle \Delta\alpha/\alpha \rangle$  
(see, e.g., Murphy et al. 2004, and references therein).

Contemporary astronomical observations still cannot provide   
accuracy comparable to the laboratory results. 
Up to now, the errors of
the fractional deviation of the fine-structure constant,\,
$\Delta\alpha/\alpha = (\alpha_z - \alpha_0)/\alpha_0$,\,
recovered from astronomical spectra
are at the level of $10^{-6}$
(or parts per million, ppm)\footnote{Here $\alpha_0$ and $\alpha_z$
denote the values of the fine-structure constant at epoch $z=0$ and
in a given absorption (or emission) line system at redshift $z$, 
respectively.}. For instance, individual 
\daa\ values obtained by Chand et al. (2004)
are known with the accuracy of a few ppm. 
The uncertainty of the same order of magnitude,\,
$\sigma_{\langle\Delta\alpha/\alpha\rangle} \simeq$ 1.7-1.9~ppm,\, 
was achieved in our previous study of the 
\zabs = 1.15 system towards \object{HE 0515--4414} 
(Quast et al. 2004, hereafter QRL). 
In both cases the standard many-multiplet (MM) method 
(Webb et al. 1999; Dzuba et al. 1999, 2002) has been used. 
Further modification of the MM method 
(Levshakov 2004; Levshakov et al. 2005, hereafter LCMD) 
resulted in a new methodology for probing the cosmological
variability of $\alpha$ on base of pairs of \ion{Fe}{ii} lines
observed in {\it individual exposures} 
from a high resolution spectrograph
(henceforth referred to as SIDAM~-- 
single ion differential $\alpha$ measurement).
The basic idea behind SIDAM
was to avoid the influence of small spectral shifts due to
ionization inhomogeneities within the absorbers and 
due to non-zero offsets between different exposures. 
The individual offsets can affect
the shape of the line profiles during rebinning and coadding procedures
which are usually applied to combine exposures together 
to increase signal-to-noise, S/N, ratio (examples are given in LCMD). 

%-----------------Figure 1
\begin{figure}[t]
\vspace{-2.5cm}
\hspace{-0.6cm}\psfig{figure=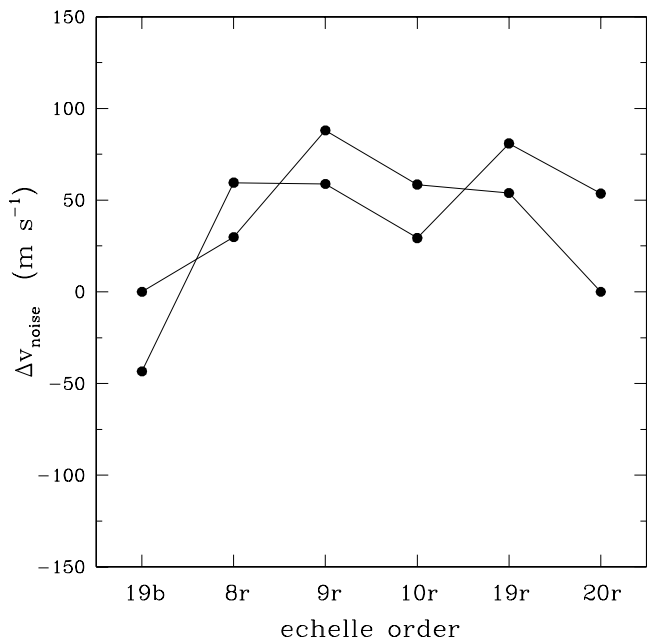,height=12.0cm,width=12.0cm}
\vspace{-1.1cm}
\caption[]{Doppler shifts of two calibrated ThAr lines 
bracketing the positions of
the \ion{Fe}{ii}  absorption complex (Fig.~2) for exposure No.1
(Table~2). Echelle orders (Table~3) are indicated on
the bottom axis (`b' and `r' denote the blue and red arms, respectively).
The ThAr lines used are the following (wavelengths in \AA):
3457.070 \& 3462.851 (19b), 5039.230 \& 5041.122 (8r),
5101.129 \& 5111.278 (9r), 5122.499 \& 5125.489 (10r),
5559.891 \& 5564.201 (19r), and 5595.063 \& 5599.654 (20r).
The peak-to-peak variations
of $\Delta v_{\rm noise} \sim 100$ \ms\, set a
lower limit to the error of the \ion{Fe}{ii}
relative position measurements
}
\label{fig1}
\end{figure}

In the present paper we show that the SIDAM can provide
a sub-ppm precision in a single redshift \daa\ measurement
and that this level of accuracy is caused 
by {\it intrinsic instrumental imperfections} and
{\it systematic errors} inherited from the uncertainties of 
the wavelength scale calibration. 

While working at sub-ppm level,
we confront the problem of {\it sub-pixel} centering.
As shown by David \& Verschueren (1995),
this problem can be properly treated only in two cases:\,
either ($i$) the exact analytical function 
describing the observed line profile is 
known a priori, or ($ii$) the observed line profile
is intrinsically symmetric.
Neither of these conditions is 
fulfilled for metal absorption lines observed in QSO spectra.
This means that in general 
the line centering must be handled with special care.     
We discuss this problem in detail in Sects.~2 and 3.
The results of the SIDAM are presented in Sect.~4.
Our conclusions are given in Sect.~5.

%-------------------- Table 2
\begin{table}[t]
\centering
\caption{ESO UVES archive data on ThAr lamp used 
to calibrate spectra of \object{HE 0515--4414} }
\label{tbl-2}
\begin{tabular}{ccccc}
\hline
\noalign{\smallskip}
ThAr&Date,&Time,&T,$^\circ$C&P, mb \\ 
No. & y-m-d & h:m & & \\
\hline
\noalign{\smallskip}
\noalign{\smallskip}
1&1999-12-14&08:17&11.8&990.0 \\
4&2000-11-17&07:58&13.7&990.1 \\
5&2000-11-18&08:41&13.3&989.1 \\
6&2000-11-19&08:44&14.1&989.3 \\
7&2000-12-25&07:29&16.5&990.0 \\
\noalign{\smallskip}
\hline
\noalign{\smallskip}
\multicolumn{5}{l}{Temperature and air pressure are taken from the}\\ 
\multicolumn{5}{l}{ESO/ST-ECF Science Archive Ambient Conditions }\\ 
\multicolumn{5}{l}{Database at http://archive.eso.org/}\\ 
\end{tabular}
\end{table}

\section{Observations and Data Reduction}

We analyze high quality
spectra of the bright intermediate redshift quasar
\object{HE 0515--4414} (\zem = 1.73, $B=15.0$; Reimers et al. 1998). 
The observations were acquired with the 
UV-Visual Echelle Spectrograph (UVES) 
at the VLT 8.2~m telescope at Paranal, Chile, and the spectral data were
retrieved from the ESO archive. 
The selected exposures are listed in Table~1.
The spectra were recorded with a dichroic filter which allows
to work with the blue and red UVES arms simultaneously as with 
two independent spectrographs, 
and the CCDs were read-out in $1\times1$ binned pixels
(spatial$\times$dispersion direction).
The standard settings at central wavelengths 
$\lambda$346 nm and $\lambda$580 nm
were used for the blue and red arms respectively (marked by symbols `b' 
and `r' in
Col.1 of Table~1 (along with the exposure number).  
From the blue spectra we used only order 19, and from 
the red spectra orders
8, 9, 10, 19 and 20\footnote{These are the sequential numbers of the
echelle orders which are used throughout the paper. The correspondence between
them and echelle orders is as follows: 19b=135, 8r=121,
9r=120, 10r=119, 19r=110, and 20r=109.}, 
where \ion{Fe}{ii} lines 
suitable for the \daa\ measurement 
are observed. All selected \ion{Fe}{ii} lines are located close 
to the central regions of the corresponding
echelle orders. This minimizes possible distortions of the line profiles
caused by the decreasing spectral
sensitivity at the edges of echelle orders (LCMD).

Other details of the observations are also presented in Table~1.
Cols. 2 and 3 give setting and slit width used.
Cols. 4 and 5 list the date and the starting time of the exposure whose 
duration is indicated in Col. 6.
The ambient conditions are characterized 
by minimum and maximum values of the seeing, 
${\cal S}$, temperature, $T$, and
air pressure, $P$ (Cols.7-9).
During an integration these quantities were  
varying between the indicated limits. 
Col. 11 shows  that
the data were obtained for two observational programs, 60.A-9022 
and 66.A-0212.
To characterize the quality of the obtained
spectra, we introduce 
a quality factor, $\kappa$ (Col. 10),
defined as\, $\kappa = 1/(\Delta {\cal S}\cdot\Delta T\cdot\Delta P)$\, 
and expressed in fractions of the 4th
exposure which has the highest quality.
The selection of variables $\Delta {\cal S},\, \Delta T$, and $\Delta P$ is 
heuristic in some sense but it is relevant to 
the problem of the line centering. 
For instance, a change of $\Delta P = 1$ millibar (or a change of $\Delta T =
0.3^\circ$C) induces an error in radial velocities of $\sim 50$ \ms
(Kaufer, D'Odorico \& Kaper 2004). 
Problems may also occur because of seeing variations 
which change the intensity
of the QSO signal during an integration, and, hence, the instabilities of the
spectrograph may be sampled in different ways.
Other effects like vibration, thermal drift, 
changes of the grating spacing due to
temperature fluctuations, unequal illumination of the spectrograph grating and
collimator by the laboratory reference source and starlight,
varying width of the instrumental profile, and scattered light 
may produce additional Doppler noise at the level of 
a few \ms\, (Griffen \& Griffen 1973; Brown 1990;
Meyer 1990; Gulliver et al. 1996). 
These deleterious effects do not permit very precise measurements of the line
positions even from extremely high S/N spectra.
However, these problems can be considerably reduced using fiber fed
spectrographs like, e.g., a configuration of UVES+FLAMES which provides
the radial velocity precision better than 100 \ms\, for the stars with
$V$ magnitudes in the range 14--18 (Melo et al. 2004; Bouchy et al. 2005).

%-------------------- Table 3
\begin{table}[t]
\centering
\caption{Original pixel sizes along the selected echelle orders}
\label{tbl-3}
\begin{tabular}{c r@{--}l c}
\hline
\noalign{\smallskip}
Echelle & \multicolumn{2}{l}{$\Delta \lambda$ (\AA) } &
$\lambda$(\ion{Fe}{ii}) complex (\AA) \\ 
order  &  \multicolumn{2}{l}{Pixel size (m\AA)} & 
Pixel size (\kms) \\
\noalign{\smallskip}
\hline
\noalign{\smallskip}
blue arm & 3449&3469 & 3459\\
\#19     & 19.5&16.0 & 1.55\\
\noalign{\smallskip}
red arm & 5030&5050 & 5040\\
\#8     & 22.7&20.5 & 1.29\\
\noalign{\smallskip}
red arm & 5095&5115 & 5104\\
\#9     & 20.4&17.9 & 1.13\\
\noalign{\smallskip}
red arm & 5113&5133 & 5123\\
\#10    & 23.2&21.0 & 1.30\\
\noalign{\smallskip}
red arm & 5552&5572 & 5562\\
\#19    & 22.9&20.5 & 1.17\\
\noalign{\smallskip}
red arm & 5581&5601 & 5591\\
\#20    & 25.3&23.2 & 1.30\\
\noalign{\smallskip}
\hline
\end{tabular}
\end{table}

We used the UVES pipeline (the routines implemented in MIDAS-ESO data
reduction package for UVES data) to perform
the bias correction, inter-order background subtraction,
flat-fielding, correction of cosmic rays impacts, 
sky subtraction, extraction of the orders in the pixel space
and wavelength calibration. 
A modified version of the routine `echred' of the context ECHELLE inside
MIDAS was used to calibrate in wavelength 
the echelle spectra {\it without rebinning}.
In this way we have the reduced spectra with their 
original pixel size in wavelength. 

The residuals of the calibrations are rather small amounting to
$\sigma_{\rm rms} \la 1$ m\AA.
The observed wavelength scale of each spectrum was transformed into vacuum,
heliocentric wavelength scale (Edl\'en 1966). 

%-------------------- Table 4
\begin{table*}[t]
\centering
\caption{The instrumental profile widths (FWHM) in \kms\, and the mean
signal-to-noise ratios per pixel (given in parentheses) 
at the continuum level at the positions
of \ion{Fe}{ii} lines in scientific exposures}
\label{tbl-4}
\begin{tabular}{cccccccc}
\hline
\noalign{\smallskip}
Echelle & \multicolumn{7}{c}{Scientific exposures}\\
order & {\#1} & {\#2} & {\#3} & {\#4} & {\#5} & {\#6} & {\#7} \\   
\noalign{\smallskip}
\hline
\noalign{\smallskip}
19b&5.60$^{+0.10}_{-0.10}$(31)&5.60$^{+0.10}_{-0.10}$(31)
&5.60$^{+0.10}_{-0.10}$(27)&5.60$^{+0.10}_{-0.10}$(31)
&5.60$^{+0.10}_{-0.10}$(31)&5.60$^{+0.10}_{-0.10}$(28)
&5.60$^{+0.10}_{-0.10}$(35)\\ 
\noalign{\smallskip}
8r&4.88$^{+0.07}_{-0.07}$(46)&4.88$^{+0.07}_{-0.07}$(50)
&4.88$^{+0.07}_{-0.07}$(43)&5.50$^{+0.10}_{-0.10}$(56)
&5.40$^{+0.10}_{-0.10}$(54)&5.50$^{+0.20}_{-0.20}$(51)
&5.50$^{+0.06}_{-0.06}$(51)\\ 
\noalign{\smallskip}
9r&4.77$^{+0.05}_{-0.05}$(51)&4.77$^{+0.05}_{-0.05}$(57)
&4.77$^{+0.05}_{-0.05}$(49)&5.36$^{+0.09}_{-0.09}$(44)
&5.43$^{+0.08}_{-0.08}$(47)&5.47$^{+0.08}_{-0.08}$(52)
&5.47$^{+0.08}_{-0.08}$(45)\\ 
\noalign{\smallskip}
10r&4.84$^{+0.09}_{-0.09}$(47)&4.84$^{+0.09}_{-0.09}$(43)
&4.84$^{+0.09}_{-0.09}$(42)&5.46$^{+0.12}_{-0.12}$(49)
&5.46$^{+0.12}_{-0.12}$(52)&5.54$^{+0.11}_{-0.11}$(50)
&5.46$^{+0.05}_{-0.05}$(53)\\ 
\noalign{\smallskip}
19r&4.78$^{+0.03}_{-0.03}$(47)&4.78$^{+0.03}_{-0.03}$(54)
&4.78$^{+0.03}_{-0.03}$(49)&5.39$^{+0.11}_{-0.11}$(53)
&5.45$^{+0.32}_{-0.32}$(54)&5.61$^{+0.27}_{-0.27}$(59)
&5.45$^{+0.11}_{-0.11}$(54)\\ 
\noalign{\smallskip}
20r&4.83$^{+0.05}_{-0.05}$(53)&4.83$^{+0.05}_{-0.05}$(49)
&4.83$^{+0.05}_{-0.05}$(45)&5.42$^{+0.05}_{-0.05}$(44)
&5.47$^{+0.11}_{-0.11}$(60)&5.53$^{+0.11}_{-0.11}$(55)
&5.42$^{+0.05}_{-0.05}$(48)\\ 
\noalign{\smallskip}
\hline
\end{tabular}
\end{table*}

Following the same procedure as
for the QSO exposures, we calibrated in wavelength the spectra of the 
ThAr arcs. This allowed us to estimate  
the `Doppler noise' 
by measuring the random velocity shifts
of the ThAr emission lines. For each order we selected two 
well-exposed ThAr emissions which
bracket the positions of the \ion{Fe}{ii} absorption complex.
In Fig.~1 we show the Doppler shifts, 
$\Delta v_{\rm noise}$,  measured with the calibrated
ThAr lamp No.1 (Table~2). 
For both lines of the arc  the peak-to-peak variations
of $\Delta v_{\rm noise}$ are about 100 \ms.
This gives us a lower limit to the error of 
\ion{Fe}{ii} positional estimations.
 
%-----------------Figure 2
\begin{figure*}[t]
\vspace{0.0cm}
\hspace{0.25cm}\psfig{figure=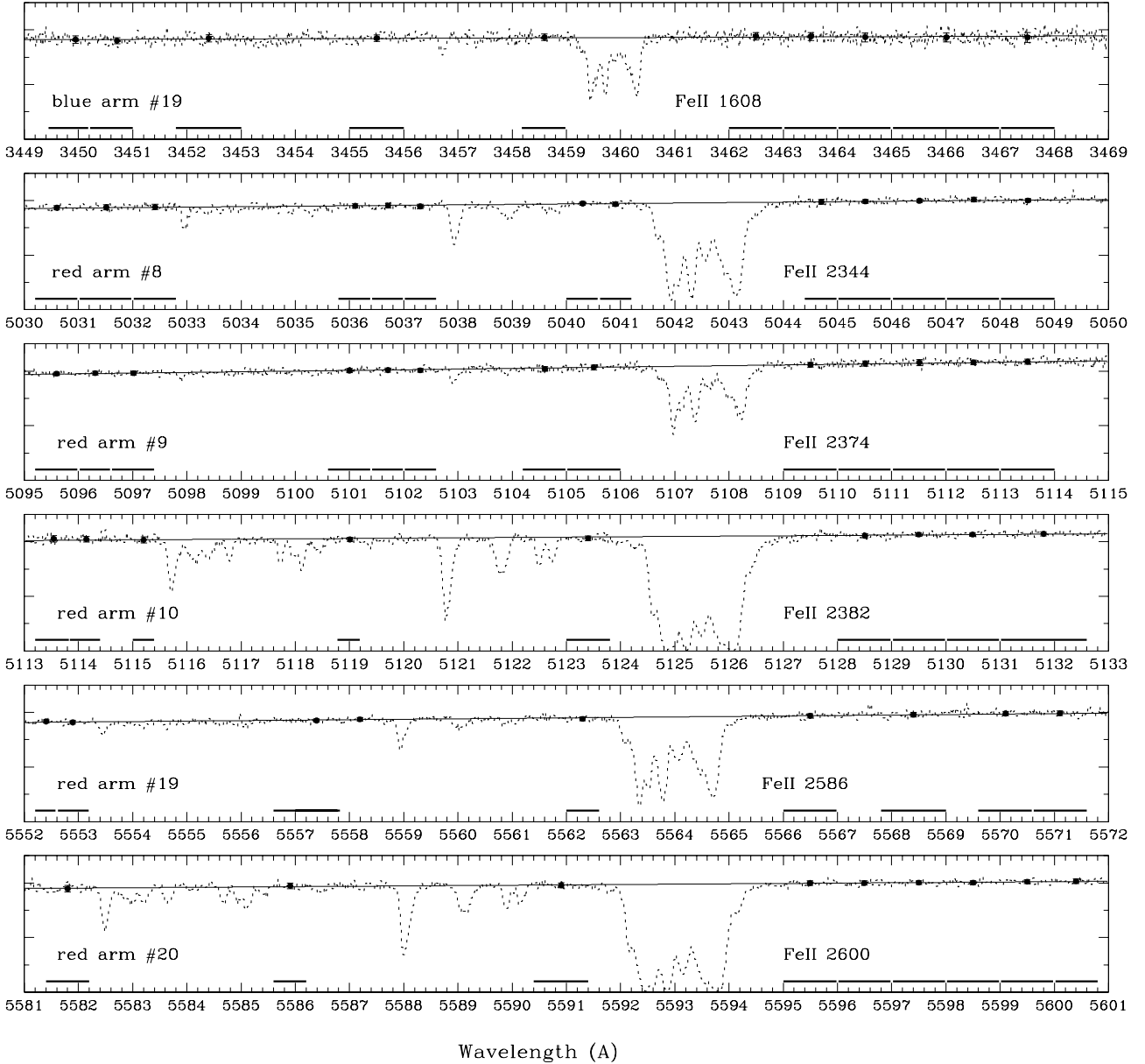,height=17.0cm,width=18.0cm}
\vspace{-1.0cm}
\caption[]{Unnormalized portions of the 
\object{HE 0515--4414} spectra (intensities are in arbitrary units)
with the \ion{Fe}{ii} lines 
(\zabs = 1.15) obtained from the 4th exposure (Table~1).
All \ion{Fe}{ii} lines are located near the center of the orders.
(indicated at the left side of each panel). 
Dots with error bars are the mean
intensities and their 1$\sigma$ errors (calculated in the ranges marked 
with horizontal lines) used to estimate the local continua  
by means of the linear regression analysis (LCMD).
The uncertainty of each local continuum level is less than 1\%
}
\label{fig2}
\end{figure*}

Usually when a precise absolute wavelength scale  
is required, calibration exposures are taken before and after the
scientific exposures.
As discussed above, varying ambient weather conditions may introduce
different velocity offsets in the lamp and QSO spectra if they were
not obtained closely in time. However, this uncertainty affects
only an absolute calibration whereas in \daa\ estimations we
are dealing with {\it differential} measurements. 
In fact, we measure the relative
\ion{Fe}{ii} positions with respect to the 
\ion{Fe}{ii} $\lambda1608$ line (LCMD). 
Since the SIDAM uses iron
lines obtained simultaneously in one exposure, a systematic offset
caused by shortcomings of the calibration procedure should be canceled 
out for the blue and red arms spectra. This is not the case if unexpected
mechanical instabilities occur during integration (see 
Sect.3 below and an example in LCMD where effects of the
mechanical instabilities are discussed).
For our data, the first three QSO exposures were calibrated with
the same ThAr lamp No.1 (Table~2). The 4th to 7th QSO
exposures were calibrated with the corresponding ThAr
lamps No.4-7 taken immediately after the scientific exposures 
(cf. Tables~1 and 2).

Our further concern was nonlinearity of the wavelength scale.
Off-plane design in echelle spectrographs like UVES 
is introducing  a pronounced curvature of the orders
(e.g. Ballester \& Rosa 1997).
As a result, the wavelength scale is not linear in the sense that the
original pixel width in wavelength decreases with increasing pixel number
along the frame.
Table~3 illustrates this behavior for the selected echelle orders.
In Col.2, upper row for each echelle order indicates the wavelength range 
which includes the \ion{Fe}{ii}
absorptions, whereas the corresponding pixel
sizes at the starting and end points of this range are given below.
Col.3 shows the pixel size at 
the center position of the \ion{Fe}{ii} complex
(lower and upper rows, respectively).
The \ion{Fe}{ii} complex is depicted in Fig.~2.

The difference between pixel sizes at the edges of the wavelength
ranges shown in Fig.~2 and Table~3
is about 18\% for order 19 (blue arm) and 8-12\%
for the red arm orders. This introduces an artificial
inclination of the local continuum level which may affect 
the relative positions of the \ion{Fe}{ii} lines 
at the sub-pixel level.
Therefore we normalized the registered
photocounts per pixel to the original pixel size
in wavelength and determined the local continua
for each \ion{Fe}{ii} complex following linear regression 
procedure from LCMD.
The uncertainties of the calculated continua 
are less than 1\%\, in all \ion{Fe}{ii} regions.
An example of the continuum fitting is shown in Fig.~2 
for the 4th QSO exposure.

Finally, we measured the FWHM of the instrumental profile 
for each individual echelle order. The instrumental profile
is dominated in our case by the slit width which was different
for the blue and red arms in the first three exposures (Table~1). 
The FWHM values calculated from the narrow lines of the arc spectra 
are given in Table~4.
Our further results were obtained with the 
mean FWHM and signal-to-noise ratio values listed in Table~4.

\section{Concordance of the \ion{Fe}{ii} profiles}

The spectra of \object{HE 0515--4414}
reveal a multi-component complex of metal absorption
lines associated with a sub-damped Ly$\alpha$ (sub-DLA) system at 
\zabs = 1.15 (de la Varga et al. 2000; 
Quast et al. 2002; Reimers et al. 2003).
The radial velocities of the
absorption components of the \ion{Fe}{ii} complex span 660 \kms\,
(QRL, Table~2). Not all of them are seen, however, in the
individual exposures, the weakest ones were detected only in the co-added
spectra (QRL, Fig.~1). 
Therefore, for our purpose we chose two sub-systems
at \zabs = 1.150965 and 1.149092 which exhibit the most pronounced
absorption features and, thus, provide the most accurate determination
of their relative positions from individual exposures.
The atomic data used for the analysis are given in Table~5.

In both sub-systems there are no isolated and symmetric absorption
components which could be used for the line centering. But, as
explained in LCMD, we are mainly interested in the position of an
absorption complex as a whole to calculate 
its radial velocity shift with respect to the
reference line \ion{Fe}{ii} $\lambda1608$~-- the only line among 
\ion{Fe}{ii} transitions which has a negative sensitivity coefficient
${\cal Q}$ (Table~5).

In general,
the position of the line blend is sensitive to the relative strengths
of the partly resolved and unresolved (hidden) components as well as to the
shape of the instrumental profile. The effect is most noticeable
in the case of optically thick and narrow hidden components 
due to their different saturation depending on the line strength
(Levshakov 1994; Levshakov \& D'Odorico 1995; QRL). 
As seen from Table~5, the strengths of the iron lines, 
$\log f\lambda_{\rm vac}$, most frequently
observed in QSO spectra (\ion{Fe}{ii} $\lambda\lambda 2374, 2382$)
differ by an order of magnitude.

To handle the blend we have to adopt a specific
model. Since we are dealing with only
one ion, all \ion{Fe}{ii}
lines from an intervening absorber {\it must} have similar profiles.
A model not matching all observational profiles simultaneously
either indicates blended lines from different intervening
systems or that some data points are outliers.
This allows us to check  
concordance of the \ion{Fe}{ii} data before evaluating \daa.

To define the model we will assume,
following LCMD, that:
($1$) the number of subcomponents $n_{\rm s}$ is fixed for 
all \ion{Fe}{ii} lines under study;
($2$) the Doppler $b_i$ parameters are identical for the same
$i$th subcomponents; ($3$) the relative intensities of the subcomponents
$r_{i,j}$ and ($4$) the relative radial velocities
$\delta v_{i,j}$ between the subcomponents are fixed.  

Then, an \ion{Fe}{ii} blend can be described by the sum of $n_{\rm s}$
Voigt functions:
\begin{equation}
\tau^{(\ell)}_v = N_1\,\sum^{n_{\rm s}}_{i=1}\,r_{i,1}\,{\cal V}\left[
(v - v_\ell - \delta v_{i,1})/{b_i} \right]\; ,
\label{eq1}
\end{equation}
where $\tau^{(\ell)}_v$ is the optical depth at radial velocity $v$ within 
the $\ell$th \ion{Fe}{ii} line ($\ell = 1,2,\ldots,L$),
$N_1$ is the column density of the main component, 
$r_{i,1} = N_i/N_1$, 
$v_\ell$ is the position of the main component in the line $\ell$,
and $\delta v_{1,1} = 0$. 
Here $L$ is the total number of \ion{Fe}{ii} lines involved in the
analysis.

%-------------------- Table 5
\begin{table}[t]
\centering
\caption{Atomic data of the \ion{Fe}{ii} transitions$^a$, and the 
sensitivity coefficients ${\cal Q}^b$. 
Estimated errors are given in parentheses}
\label{tbl-5}
\begin{tabular}{c r@{.}l r@{.}l r@{.}l c}
\hline
\noalign{\smallskip}
Mlt. & \multicolumn{2}{c}{$\lambda_{\rm vac}$, } &
\multicolumn{2}{c}{$f$} & \multicolumn{2}{c}{\hspace{-0.5cm}${\cal Q}$} &
$\log f\lambda_{\rm vac}$ \\ 
No.$^c$  &  \multicolumn{2}{c}{\AA} & \multicolumn{2}{c}{ } & 
\multicolumn{2}{c}{ } & \\
\noalign{\smallskip}
\hline
\noalign{\smallskip}
1u & 2600&1725(1) & 0&23878 & 0&035(4) & 2.79\\
1u & 2586&6496(1) & 0&06918 & 0&039(4) & 2.25\\
2u & 2382&7642(1) & 0&320   & 0&035(4) & 2.88\\
2u & 2374&4603(1) & 0&0313  & 0&038(4) & 1.87\\
3u & 2344&2130(1) & 0&114   & 0&028(4) & 2.43\\
8u & 1608&45080(8)& 0&0580  & --0&021(5)& 1.97\\
\noalign{\smallskip}
\hline
\noalign{\smallskip}
\multicolumn{8}{l}{$^a$based on the compilation of Murphy et al. (2003);}\\ 
\multicolumn{8}{l}{$^b$defined in Sect.~4; 
$^c$multiplet numbers from Morton (2003)}
\end{tabular}
\end{table}

%-----------------Figure 3
\begin{figure}[t]
\vspace{-2.5cm}
\hspace{-0.35cm}\psfig{figure=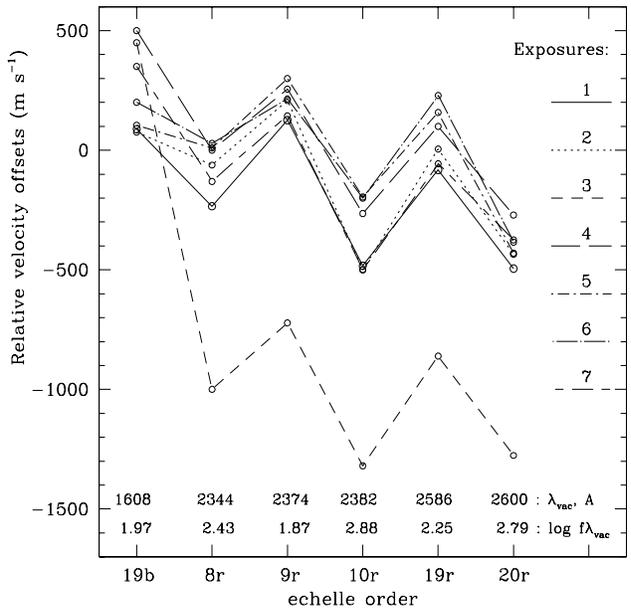,height=12.0cm,width=11.5cm}
\vspace{-1.3cm}
\caption[]{
Same as Fig.~1 but for the \ion{Fe}{ii} lines from the
\zabs = 1.150965 system.
The exposure numbering is given in accord with Table~1, whereas
the line strengths and wavelengths~-- according to Table~5.
The relative velocity offsets are based on the 8-component model
estimations (see text for details)
}
\label{fig3}
\end{figure}

The model is fully defined by specifying 
$p = 3n_{\rm s}+L-1$ parameters:  
$N_1$, 
$\{b_i\}^{n_{\rm s}}_{i=1}$,
$\{\Delta v_{i,1}\}^{n_{\rm s}-1}_{i=1}$,
$\{r_{i,1}\}^{n_{\rm s}-1}_{i=1}$, and
$\{v_\ell\}^{L}_{\ell=1}$.
All these parameters are components of the parameter vector
$\theta = \{\theta_1, \theta_2, \ldots, \theta_p\}$.
To estimate $\theta$ from the 
combined \ion{Fe}{ii} profiles, we minimize the objective
function
\begin{equation}
\chi^2(\theta) = \frac{1}{\nu}\,\sum^{L}_{\ell=1}\,\sum^{m_\ell}_{j=1}\,
\left[ {\cal F}^{\rm cal}_{\ell,j}(\theta) - {\cal F}^{\rm obs}_{\ell,j}
\right]^2/\sigma^2_{\ell,j}\; ,
\label{eq2}
\end{equation}
where ${\cal F}^{\rm obs}_{\ell,j}$ is the observed normalized intensity of the
spectral line $\ell$, $\sigma_{\ell, j}$ is the experimental error within the
$j$th pixel of the line profile, 
and $\nu = \sum^L_\ell\,m_\ell\, -\,p\, \equiv M-p$ 
is the number of degrees of freedom.
${\cal F}^{\rm cal}_{\ell,j}(\theta)$ is the
calculated intensity convolved with the 
corresponding spectrograph point-spread function (Table~4).

%-----------------Figure 4
\begin{figure}[t]
\vspace{0.0cm}
\hspace{-0.3cm}\psfig{figure=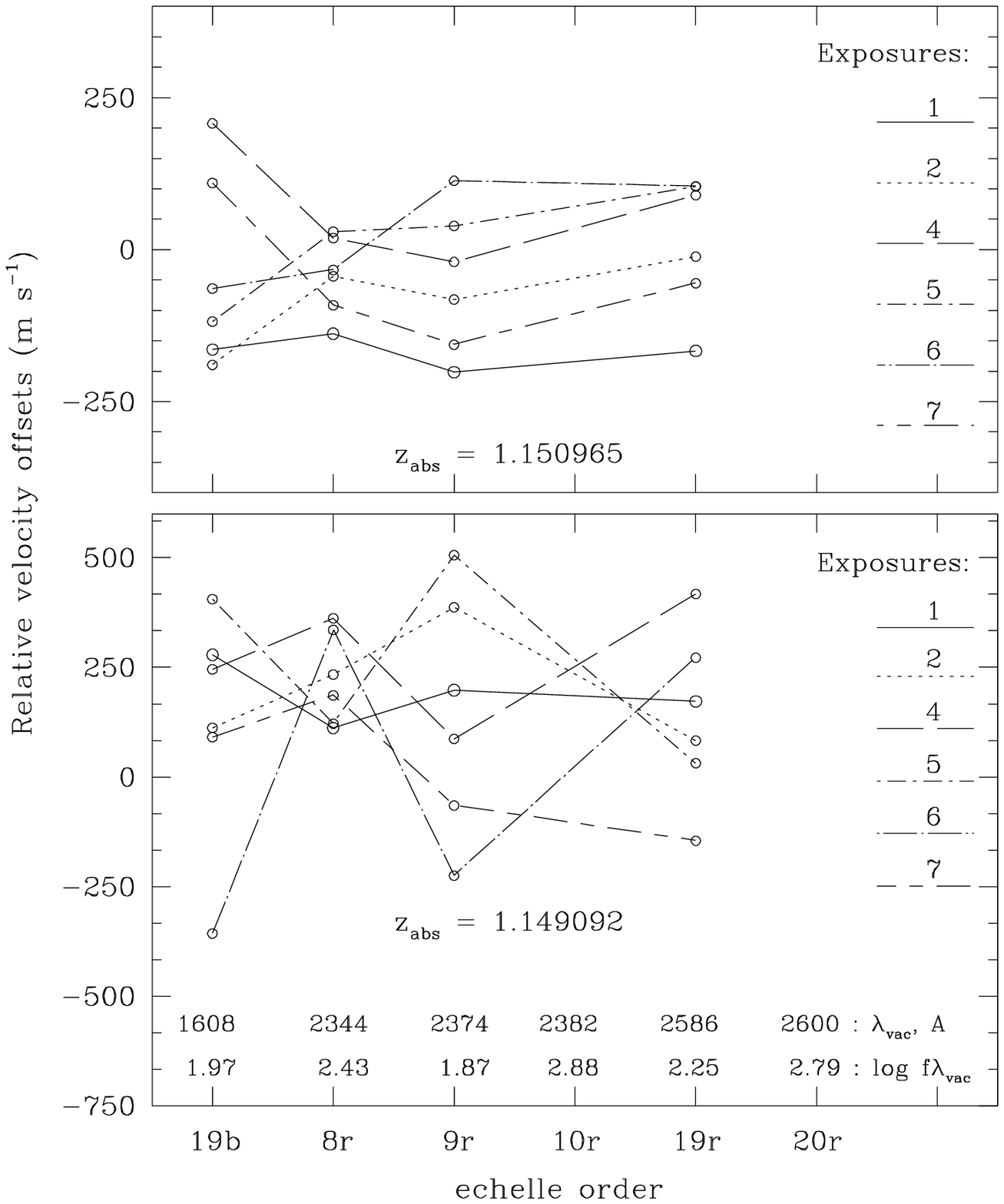,height=12.0cm,width=11.5cm}
\vspace{-1.7cm}
\caption[]{
Same as Fig.~3 but for the \ion{Fe}{ii} lines from the
\zabs = 1.150965 and \zabs = 1.149092 systems.
The corresponding models are 
superpositions of $n_{\rm s} = 13$ and $n_{\rm s} = 2$ Voigt profiles.
The upper panel illustrates the effect of hidden (unresolved)
components: the peak-to-peak variations
of $\Delta v$ are decreasing with increasing number of sub-components.
Smaller equivalent widths of the \ion{Fe}{ii} lines at \zabs = 1.149092
lead to a larger dispersion of the $\Delta v$ values shown in the lower
panel
}
\label{fig4}
\end{figure}

The total number of data points involved in the analysis is
$M=2704$ (\zabs = 1.150965) and $M=428$ ( \zabs = 1.149092). 
All selected \ion{Fe}{ii} profiles are free 
from cosmic rays and telluric absorptions.
In accord with QRL, we took $n_{\rm s} = 8\; (p=65)$ and 
$n_{\rm s} = 2\; (p=47)$
to define our models for the former and the latter sub-systems.
We found that the models chosen described both sets of data adequately:
the $\chi^2_\nu$
per degree of freedom was about 1 in both cases. 
Thus, each ensemble of \ion{Fe}{ii} lines 
is self-consistent, free from outliers and blending with
other lines.

%-----------------Figure 5
\begin{figure*}[t]
\vspace{0.0cm}
\hspace{0.0cm}\psfig{figure=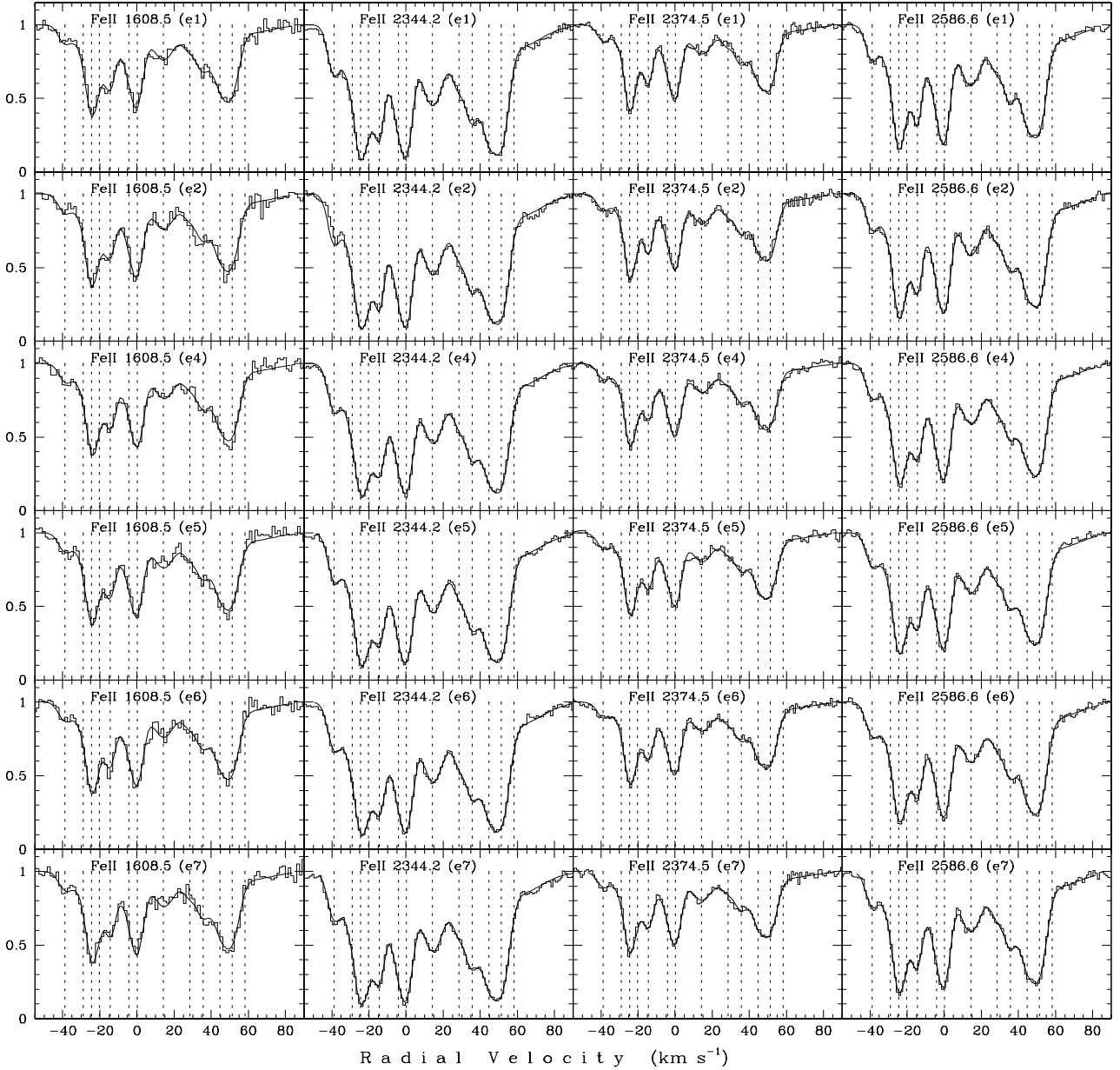,height=17.0cm,width=18.0cm}
\vspace{-0.7cm}
\caption[]{Individual exposures (labeled as `e\#')
of the \ion{Fe}{ii} lines 
selected from the spectrum of \object{HE 0515--4414}
to estimate \daa. 
Normalized intensities are indicated by histograms. 
The over-potted synthetic profiles (smooth curves) are 
calculated from the joint analysis of all iron lines.
The zero radial velocity is fixed at $z = 1.150965$.
The minimization procedure gives $\chi^2_{\rm min} = 0.9$ per degree of
freedom ($\nu = 2642$).
The dashed vertical lines mark positions of the 
sub-components from Table~6
}
\label{fig5}
\end{figure*}

This preliminary analysis revealed, however, an unexpectedly large
shift between the blue and red arms for the 3rd exposure
(short-dashed line in Fig.~3). It might be a chance coincidence,
but the spectral data obtained from the 3rd exposure 
have the lowest quality factor $\kappa = 0.006$ as compared with unity
(see Table~1). The quality factors for other exposures are
much larger and do not exhibit significant variations in their values. 
Similar offsets between the blue and red arms 
were noted for the UVES spectra
of \object{Q 1101--264} in LCMD where we ascribed them
to mechanical instabilities. Such exposures 
were excluded from further
\daa\ measurements since they may induce a mock signal.

%-----------------Figure 6
\begin{figure*}[t]
\vspace{0.0cm}
\hspace{0.0cm}\psfig{figure=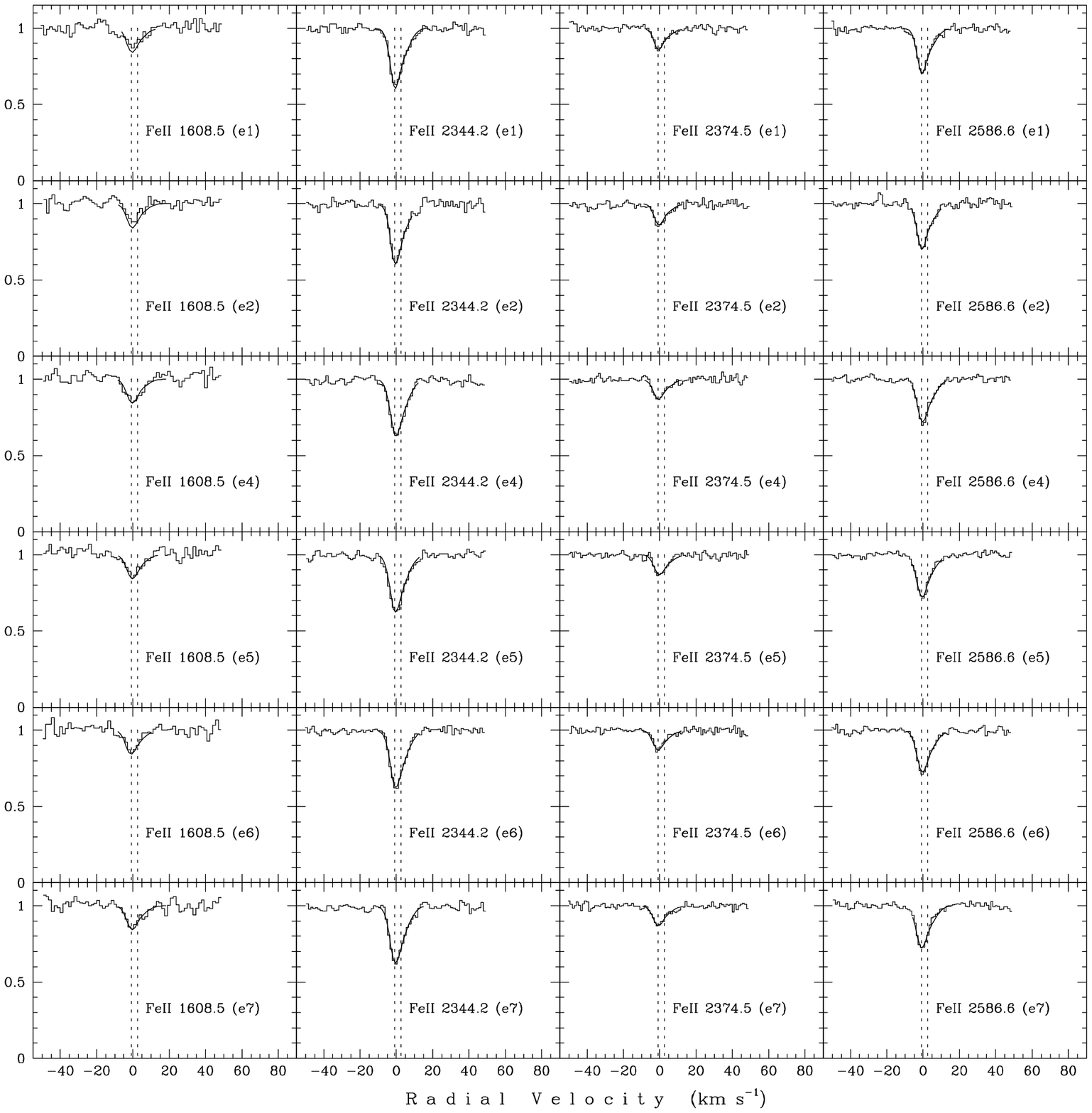,height=17.0cm,width=18.0cm}
\vspace{-0.7cm}
\caption[]{Same as Fig.~5 but for the sub-system at \zabs = 1.149092.
The minimization procedure gives $\chi^2_{\rm min} = 0.8$ per degree of
freedom ($\nu = 399$).
The dashed vertical lines mark positions of the 
sub-components from Table~7
}
\label{fig6}
\end{figure*}

Fig.~3 shows one more problem: there are too large
peak-to-peak variations of the relative velocity offsets,
$\Delta v$, and we note a clear dependence of $\Delta v$ on the line strength. 
This behavior can be corrected by excluding saturated lines (like 
\ion{Fe}{ii} $\lambda2382$ and $\lambda2600$) from the analysis,
and by increasing the number of sub-components to account
for the influence of the unresolved (hidden) blends
on the line centering. 
The number of sub-components is controlled by two factors: 
($i$) the value of $\chi^2_{\rm min}$ should be
close to unity, while ($ii$) the dispersion of $\Delta v$  
should decrease but not differ much from that deduced
from ThAr lines shown in Fig.~1. 
Fig.~4 (upper panel) shows the results obtained for
a model with $n_{\rm s} = 13$ sub-components ($p = 62$). 

The estimation of the uncertainties of the best-fitting model parameters
in case of many-dimensional parameter space requires a special approach.
When $p \gg 1$, the parameters are as a rule correlated, 
the parameter space near the minimum of the objective function
has a complicated topology
and the $\Delta \chi^2$ levels for the confidence regions cannot
be accurately defined. In such cases Monte Carlo simulations
are recommended instead of the formal inversion of the covariance matrix
(see, e.g., Press et al. 1989, Chapter~15).
We used the bootstrapping residuals method
to estimate the standard error ($\sigma_{\rm rms}$) of the best-fitting
parameters. The trial datasets are created in the following way.
Suppose $\tilde{\theta}$ is the estimated parameter vector and
$r_{\ell,j} = {\cal F}^{\rm cal}_{\ell,j} - {\cal F}^{\rm obs}_{\ell,j}$ 
are the residuals. 
Then a new dataset ${\cal F}^{\rm obs^\ast}_{\ell,j}$ 
is made by sampling independently from $r_{\ell,j}$ 
for each line $\ell$, yielding $r^\ast_{\ell,j}$,
and by setting 
${\cal F}^{\rm obs^\ast}_{\ell,j} =  {\cal F}^{\rm cal}_{\ell,j}(\tilde{\theta}) +
r^\ast_{\ell,j}$.
New values $\tilde{\theta}^\ast$ of the parameter vector
are then computed from the bootstrap
data in the same way that $\tilde{\theta}$ was computed from the original data,
i.e., by least squares.

%-------------------- Table 6
\begin{table*}[t]
\centering
\caption{SIDAM analysis: 
relative positions of the \ion{Fe}{ii} lines
(with respect to the adopted redshift of 1.150965),
$\delta v$, the line broadening velocities, $b$, the column
densities, $\log N$, and the corresponding rms errors (given
in parentheses) of the 13 components constituting 
the \ion{Fe}{ii} absorption complex at \zabs = 1.151
}
\label{tbl-6}
\begin{tabular}{ccccccc}
\hline
\noalign{\smallskip}
 & \multicolumn{6}{c}{Positions of the main subcomponent in} \\
\multicolumn{1}{l}{\ion{Fe}{ii}}&
\multicolumn{6}{c}{different exposures, $v$\, (\kms)}\\ 
\multicolumn{1}{l}{line (\AA)}&\#1&\#2&\#4&\#5&\#6&\#7 \\
\noalign{\smallskip}
\hline
\noalign{\smallskip}
1608& 0.536(0.086)& 0.511(0.085)& 0.907(0.086)& 0.582(0.085)& 0.636(0.086)& 
0.810(0.087) \\
2344& 0.533(0.076)& 0.688(0.076)& 0.790(0.077)& 0.804(0.077)& 0.804(0.077)& 
0.645(0.077) \\
2374& 0.562(0.087)& 0.656(0.084)& 0.719(0.086)& 0.729(0.083)& 0.667(0.085)& 
0.609(0.085) \\
2586& 0.498(0.080)& 0.618(0.080)& 0.679(0.080)& 0.739(0.080)& 0.813(0.080)& 
0.544(0.079) \\
\noalign{\medskip}
 &\multicolumn{6}{c}{Model parameters}\\
Component&\multicolumn{2}{c}{$\delta v$,}&\multicolumn{2}{c}{$b$,}&
\multicolumn{2}{c}{$\log N$,}\\
No.&\multicolumn{2}{c}{\kms}&\multicolumn{2}{c}{\kms}&\multicolumn{2}{c}{\cm}\\
\noalign{\smallskip}
\hline
\noalign{\smallskip}
1&\multicolumn{2}{c}{0.000}&\multicolumn{2}{c}{2.55(0.13)}&
\multicolumn{2}{c}{13.243(0.049)}\\
2&\multicolumn{2}{c}{-39.23(0.10)}&\multicolumn{2}{c}{3.71(0.08)}&
\multicolumn{2}{c}{12.701(0.080)}\\
3&\multicolumn{2}{c}{-30.02(0.67)}&\multicolumn{2}{c}{4.07(0.76)}&
\multicolumn{2}{c}{12.790(0.079)}\\
4&\multicolumn{2}{c}{-24.96(0.19)}&\multicolumn{2}{c}{2.31(0.09)}&
\multicolumn{2}{c}{13.458(0.075)}\\
5&\multicolumn{2}{c}{-20.56(0.40)}&\multicolumn{2}{c}{2.01(0.33)}&
\multicolumn{2}{c}{12.859(0.099)}\\
6&\multicolumn{2}{c}{-14.95(0.17)}&\multicolumn{2}{c}{2.33(0.17)}&
\multicolumn{2}{c}{13.245(0.081)}\\
7&\multicolumn{2}{c}{-3.85(0.37)}&\multicolumn{2}{c}{5.05(0.15)}&
\multicolumn{2}{c}{13.268(0.111)}\\
8&\multicolumn{2}{c}{13.74(0.09)}&\multicolumn{2}{c}{7.33(0.12)}&
\multicolumn{2}{c}{13.205(0.068)}\\
9&\multicolumn{2}{c}{28.46(0.37)}&\multicolumn{2}{c}{4.69(0.49)}&
\multicolumn{2}{c}{12.893(0.111)}\\
10&\multicolumn{2}{c}{35.23(0.14)}&\multicolumn{2}{c}{3.03(0.31)}&
\multicolumn{2}{c}{12.987(0.102)}\\
11&\multicolumn{2}{c}{44.04(0.20)}&\multicolumn{2}{c}{4.35(0.25)}&
\multicolumn{2}{c}{13.287(0.078)}\\
12&\multicolumn{2}{c}{50.68(0.14)}&\multicolumn{2}{c}{3.96(0.10)}&
\multicolumn{2}{c}{13.313(0.080)}\\
13&\multicolumn{2}{c}{58.12(1.22)}&\multicolumn{2}{c}{18.53(0.93)}&
\multicolumn{2}{c}{12.913(0.054)}\\
\noalign{\smallskip}
\hline
\end{tabular}
\end{table*}

%-------------------- Table 7
\begin{table*}[t]
\centering
\caption{Same as Table~6 but for the \ion{Fe}{ii} absorption complex 
at \zabs = 1.149092 (the two component model)
}
\label{tbl-7}
\begin{tabular}{ccccccc}
\hline
\noalign{\smallskip}
 & \multicolumn{6}{c}{Positions of the main subcomponent in} \\
\multicolumn{1}{l}{\ion{Fe}{ii}}&
\multicolumn{6}{c}{different exposures, $v$\, (\kms)}\\ 
\multicolumn{1}{l}{line (\AA)}&\#1&\#2&\#4&\#5&\#6&\#7 \\
\noalign{\smallskip}
\hline
\noalign{\smallskip}
1608& -0.821(0.557)& -0.988(0.572)& -0.854(0.413)& -0.695(0.376)& -1.457(0.319)& 
-1.010(0.226) \\
2344& -0.989(0.112)& -0.867(0.096)& -0.739(0.104)& -0.978(0.160)& -0.762(0.121)& 
-0.914(0.152) \\
2374& -0.902(0.109)& -0.713(0.245)& -1.014(0.212)& -0.595(0.198)& -1.325(0.255)& 
-1.165(0.119) \\
2586& -0.928(0.088)& -1.017(0.069)& -0.683(0.183)& -1.068(0.167)& -0.828(0.106)& 
-1.245(0.106) \\
\noalign{\medskip}
 &\multicolumn{6}{c}{Model parameters}\\
Component&\multicolumn{2}{c}{$\delta v$,}&\multicolumn{2}{c}{$b$,}&
\multicolumn{2}{c}{$\log N$,}\\
No.&\multicolumn{2}{c}{\kms}&\multicolumn{2}{c}{\kms}&\multicolumn{2}{c}{\cm}\\
\noalign{\smallskip}
\hline
\noalign{\smallskip}
1&\multicolumn{2}{c}{0.000}&\multicolumn{2}{c}{1.51(0.01)}&
\multicolumn{2}{c}{12.501(0.002)}\\
2&\multicolumn{2}{c}{3.39(0.02)}&\multicolumn{2}{c}{5.60(0.01)}&
\multicolumn{2}{c}{12.536(0.002)}\\
\noalign{\smallskip}
\hline
\end{tabular}
\end{table*}

The model parameters and their uncertainties computed on the
base of 30 bootstrap samples are given in Table~6. 
The peak-to-peak variations are now diminished by 
a factor of 2-3 and are equal to 200 \ms in average.

The lower panel in Fig.~4 shows the peak-to-peak variations for the second
sub-system with \zabs = 1.149092. A larger dispersion of $\Delta v$
in this case (peak-to-peak variations $\simeq 300$ \ms, except
the 6th exposure) is a consequence of the lower accuracy of the line
centering because of considerably smaller equivalent widths of the 
\ion{Fe}{ii} lines. Table~7 lists the model parameters for 
the second \ion{Fe}{ii} ensemble where $n_{\rm s} = 2$ ($p = 29$).   

Fig.~5 (\zabs = 1.150965) and Fig.~6 
(\zabs = 1.149092) show
the synthetic profiles (smooth curves) 
with corresponding 
QSO data (normalized intensities) plotted by histograms. 
The positions of the sub-components are marked by vertical dotted lines.
The corresponding $\chi^2_{\rm min}$ values are 0.9 ($\nu = 2642$)
and 0.8 ($\nu = 399$) for the former and the latter
sub-systems, respectively.

%-------------------- Table 8
\begin{table*}[t]
\centering
\caption{SIDAM analysis: \daa\ (ppm)
calculated with eq.(\ref{eq3}) 
and their $\sigma_{\rm rms}$ errors (given in parentheses) 
}
\label{tbl-8}
\begin{tabular}{crrr}
\hline
\noalign{\smallskip}
Exp. & \multicolumn{3}{c}{\ion{Fe}{ii} pairs}  \\
No. & 
\multicolumn{1}{c}{1608/2344} & 
\multicolumn{1}{c}{1608/2374} & 
\multicolumn{1}{c}{1608/2586} \\ 
\noalign{\smallskip}
\hline
\noalign{\smallskip}
\multicolumn{4}{c}{\zabs = 1.150965} \\
\noalign{\smallskip}
1 & 0.07(3.92) & -0.73(3.45) & 1.03(3.27) \\
2 & -6.05(3.90) & -4.12(3.38) & -2.97(3.26) \\
4 & 3.99(3.93) & 5.32(3.44) & 6.34(3.26)  \\
5 & -7.57(3.90) & -4.17(3.37) & -4.36(3.24) \\
6 & -5.74(3.92) & -0.88(3.41) & -4.93(3.26) \\
7 & 5.61(3.95) & 5.68(3.42) & 7.39(3.27) \\
\noalign{\medskip}
\multicolumn{4}{c}{\zabs = 1.149092} \\
\noalign{\smallskip}
1 & 5.69(19.34) & 2.28(16.05) & 2.95(15.68) \\
2 & -4.12(19.74) & -7.77(17.59) & 0.83(16.02) \\
4 & -3.93(14.51) & 4.51(13.14) & -4.76(12.57) \\
5 & 9.64(13.91) & -2.84(12.02) & 10.37(11.44) \\
6 & -23.57(11.61) & -3.73(11.54) & -17.47(9.34) \\
7 & -3.27(9.27) & 4.40(7.23) & 6.55(6.95) \\
\noalign{\smallskip}
\hline
\end{tabular}
\end{table*}

\section{\daa\ measurements}

The relativistic corrections of the \ion{Fe}{ii} transition
frequencies to the changes in $\alpha$ (the so-called $q$-coefficients)
have been calculated by Dzuba et al. (2002). 
In our analysis we use the dimensionless sensitivity coefficients,
${\cal Q} \equiv q/\omega_0$, listed in Table~5 (here 
$\omega_0 = 1/\lambda_0$ is the laboratory wavenumber).
Then, the value of $\Delta\alpha/\alpha$
can be estimated from a pair of lines with different
sensitivity coefficients. 
In linear approximation ($|\Delta\alpha/\alpha|\ll1$), 
eq.(5) from LCMD can be re-written in the form:
\begin{equation}
\frac{\Delta\alpha}{\alpha} = \frac{(v_2 - v_1)}
{2\,c\,({\cal Q}_1 - {\cal Q}_2)}\; ,
\label{eq3}
\end{equation}
where the line positions $v_1$ and $v_2$ are taken from the
same exposure. Here index `1' is assigned to the line $\lambda1608$, while
index `2' marks one of the other \ion{Fe}{ii} lines ($\lambda2344$, 
$\lambda2374$, or $\lambda2586$). 
The calculated \daa\ values are given in Table~8 
along with their rms errors estimated by the standard method of
error propagation.  

We now have everything we need to compute statistics:
the mean value $\langle \Delta\alpha/\alpha\rangle$ and its error. 
We note that for each individual exposure
the uncertainty of \daa\ is dominated by the error of the line centering
of the `blue' \ion{Fe}{ii} $\lambda1608$ line since its strength
is smaller as compared with the `red' lines and the \ion{Fe}{ii} $\lambda1608$
spectra have systematically lower S/N. 
Our approach reveals also a systematic effect for all the
exposures except the first one for the subsystem \zabs = 1.150965: 
the individual values of \daa\ have the same sign. 
At \zabs = 1.149092, the sign of \daa\ does not vary in 
the 1st and 6th exposures, whereas different sings are observed in
the 4th, 5th, and 7th exposures. 
However, being combined the whole sample provides
19 negative and 17 positive \daa\ values which randomize these systematic
shifts. Moreover, from inspecting the data listed in Table~8,
one finds that two values $-23.569$ and $-17.466$ from the 6th exposure 
(\zabs = 1.149092) are sharply distinct from the others and exceed 
$3\sigma_{\rm rms}$ from the weighted mean. These measurements
were rejected from the further analysis. Thus the final ensemble  
consists of the equal number ($n = 17$) 
of negative and positive \daa\ values.

For the weights, $w_i$, we use $\sigma^{-2}_{\rm rms}$ values from Table~8.
The unbiased estimators of the weighted mean 
$\tilde{x} = \langle \Delta\alpha/\alpha \rangle$, and the variance of
$\tilde{x}$ are the quantities (Linnik 1961): 
\begin{equation}
\tilde{x} = \frac{[w\,x]}{[w]}\; ,
\label{eq4}
\end{equation}
and
\begin{equation}
Var(\tilde{x}) = \frac{[w\,\tilde{a}\,\tilde{a}]}{(n-1)[w]}\; ,
\label{eq6}
\end{equation}
where $\tilde{a} = x_i - \tilde{x}$ are `apparent errors', and square
brackets denote summation, e.g., 
$[w] = \sum^n_{i=1}\,w_i$,
$[wx] = \sum^n_{i=1}\,w_i\,x_i$.

We find that the weighted mean of the ensemble of $n = 34$ \daa\ values
and the accuracy of its determination are equal to 
$\langle \Delta\alpha/\alpha \rangle = -0.07\pm0.84$ ppm ($1\sigma$ C.L.).

\section{Discussion and Conclusions}

We re-analyzed the profiles of the \ion{Fe}{ii} lines
associated with the sub-DLA system observed at \zabs = 1.15 in the
spectrum of \object{HE 0515--4414}. 
Our main purpose was to achieve the highest possible accuracy in
the line centering to set the most stringent constraint at a single
redshift to the hypothetical variability of the fine-structure constant.

This work was inspired by our recent results (LCMD)
showing the influence of reduction errors and instabilities in the
instrument on the relative radial velocity shifts in the merged QSO
spectra. It was demonstrated that co-added spectra, minded to enhance the
signal-to-noise ratio, may affect the shapes of narrow absorption lines
because each calibrated exposure has its own velocity offset.

In the previous estimations of \daa\ from the co-added
spectrum of \object{HE 0515--4414} (QRL), 
we found that in spite of a good fit of the 8-component
model of the \ion{Fe}{ii} profiles to the high S/N data,
the normalized residuals showed an unexpected
pattern in their radial velocity distribution (see Fig.~7, upper
panel). It was suggested that this pattern may be caused by
the presence of unresolved narrow lines
(or non-Gaussian line profiles) which
may induce some kind of correlations between the residuals. We now
recognize that additional correlations may be due to 
zero-point errors as well.

If individual scientific exposures are treated separately
and the number of the model components is increased up to 13, 
the pattern is
vanished and the residuals become uncorrelated (see Fig.~7, lower panel).
This demonstrates 
a clear advantage of the differential measurements. Moreover,
the precision of the new estimate of
$\langle \Delta\alpha/\alpha \rangle$ is improved by a factor of 2.
From a general point of view, such an improvement would require 4 times
longer total exposure time under the same observational conditions
(i.e., 36h vs 9h, in our case).

%-----------------Figure 7
\begin{figure}[t]
\vspace{-0.6cm}
\hspace{-0.5cm}\psfig{figure=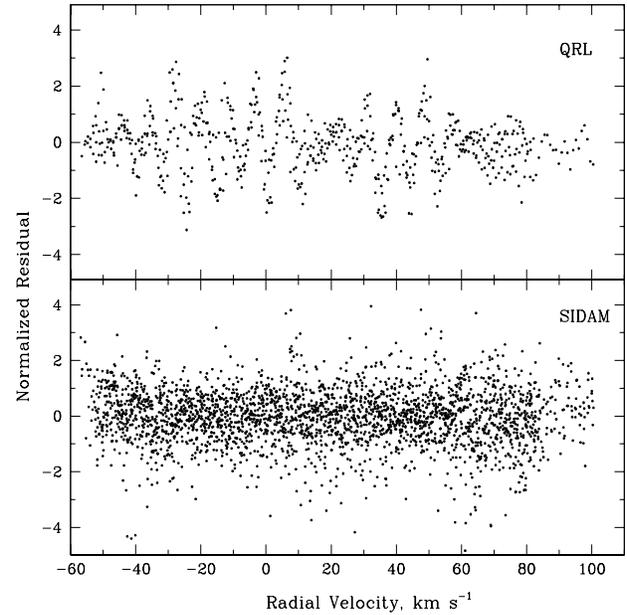,height=12.0cm,width=11.5cm}
\vspace{-3.0cm}
\caption[]{The normalized residuals,
$({\cal F}^{cal}-{\cal F}^{obs})/\sigma$, 
for the \ion{Fe}{ii} profiles
($\lambda\lambda1608, 2344, 2374$, and 2586 \AA) calculated
from the 8-component model 
in QRL ({\it upper panel}) and
from the 13-component model 
in the present work 
({\it lower panel}). 
The zero radial velocity is fixed at $z = 1.150965$
}
\label{fig7}
\end{figure}

In this respect, the present work indicates that all steps in the
data reduction procedure and in the long-term stability of the
instrument must be of a particular concern while dealing with
sub-pixel positional measurements with echelle spectrographs.
To make full and accurate utilization of the information derived
from the observations a detailed knowledge of the instrumental
characteristics is required.

In the present analysis, we have
reached an accuracy in the line centering 
which is comparable to the accuracy of the wavelength scale
calibration, $\sigma_{\rm rms} \sim 1$ m\AA. 
From eq.(\ref{eq3}) it follows that this value corresponds to
the systematic error of the individual determination of \daa\
of $\sigma_{\rm sys} \sim 4$~ppm (calculated
from a pair of \ion{Fe}{ii} lines
with $|\Delta{\cal Q}| = 0.06$).
This adds a systematic error of about 1~ppm to the total error
budget of the weighted mean,
$\sigma^2_{\rm tot} = \sigma^2_{\langle \Delta\alpha/\alpha \rangle}
+ \sigma^2_{\rm sys} \simeq 1.19$ ppm$^2$. 

It should be emphasized that
the level of $\langle \Delta\alpha/\alpha \rangle~=
(-5.7\pm1.1)$ ppm found by Murphy et al. (2004) was estimated from a
large sample of 143 absorption systems ranging from $z=0.2$ to 4.2
which were observed with the HIRES/Keck spectrograph
(here the error of the mean includes only statistical uncertainties).
Since for the individual system at $z=1.15$ the accuracy of
$\langle \Delta\alpha/\alpha \rangle$ is better than 1~ppm, we are
now in the position of arguing if the Keck ensemble average is biased.

%-----------------Figure 8
\begin{figure}[t]
\vspace{-1.5cm}
\hspace{-0.3cm}\psfig{figure=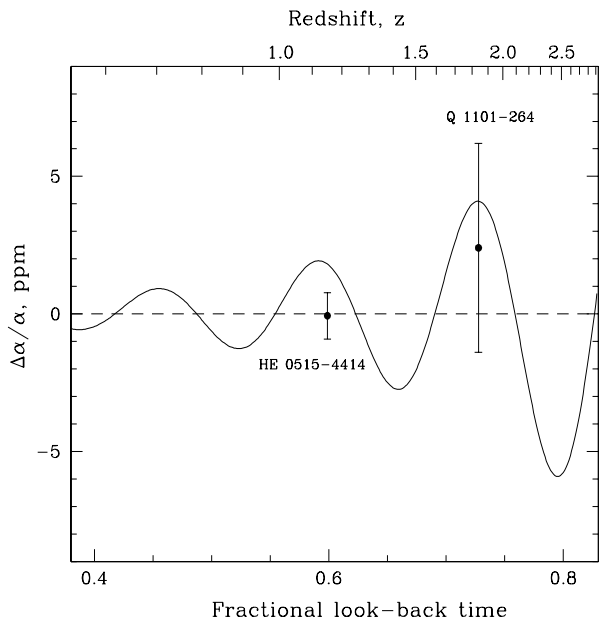,height=12.0cm,width=11.5cm}
\vspace{-2.3cm}
\caption[]{The curve illustrating an oscillatory 
behavior of \daa\ as a function of time is taken from
the middle panel of Fig.~1 in Fujii (2005).
The SIDAM results shown by dots with $1\sigma$ error bars 
(indicated are statistical errors) are from the
\zabs=1.15 system towards \object{HE 0515--4414} (present work)
and from the \zabs=1.839 system towards \object{Q 1101--264} (LCMD).
The fractional look-back time is calculated with the cosmological
parameters $h=0.7$, $\Omega_M=0.3$, and $\Omega_\Lambda=0.7$ 
}
\label{fig8}
\end{figure}

We notice that
the level of \daa\ expected at $z=1.15$ from the damped-oscillatory
model by Fujii (2005) is 2~ppm (see Fig.~8).
We do not see this value
in our analysis. However, the total error of
$\langle \Delta\alpha/\alpha \rangle$ is not small enough to
verify or reject Fujii's model.
With the probability level of 0.05 we cannot take the observed
difference between $-0.07$ ppm and 2 ppm
as significant according to the $t$-test.
To probe the oscillatory behavior of $\alpha$, 
very accurate measurements of \daa\ at higher redshifts (where the
amplitude of \daa\ is expected to be $\pm5$~ppm) are required. 

As a conclusion, it is worthwhile to note
that the achieved accuracy of \daa\ is unique for the standard
UVES configuration and that further improvement 
at the sub-ppm level can be attained only with
increasing spectral resolution and stabilizing instrumental performance 
such as, for instance, 
a fiber link producing a stable illumination at the entrance
of the spectrograph and allowing continuous 
simultaneous comparison spectrum.

\begin{acknowledgements}
S.A.L. gratefully acknowledges the hospitality 
of Hamburger Sternwarte where this work was completed.
This research has been supported by
the RFBR Grant No. 03-02-17522 and by the RLSS Grant No. 1115.2003.2,
and by the Deutsche Forschungsgemeinschaft under Grant No.
Re~353/48-1.
\end{acknowledgements}

\end{document}